\documentstyle[preprint,aps]{revtex}

\newcommand{\sigome}{$\sigma$-$\omega$ }

\begin{document}

\preprint{}

%\draft

\title{General Relativistic Mean Field Theory for Rotating Nuclei}

\author{Hideki Madokoro$^{1}$ and Masayuki Matsuzaki$^{2}$}
\address{$^{1}$Department of Physics, Kyushu University,
Fukuoka 812-81, Japan \\ $^{2}$Department of Physics, Fukuoka University
of Education, Munakata, Fukuoka 811-41, Japan}
\date{\today}
\maketitle

\begin{abstract}
We formulate a general relativistic mean field theory for rotating
nuclei starting from the special relativistic \sigome model
Lagrangian. The tetrad formalism is adopted to generalize the model
to the accelerated frame.
\end{abstract}

\pacs{PACS number: 21.60.-n, 21.30.Fe, 24.10.Jv}

Quantum hadrodynamics(QHD) is a quantum field theory which treats
nucleons and mesons as elementary degrees of freedom.
The origin of relativistic nuclear models can
be traced back to the work of Duerr\cite{ref:Du56} who reformulated a
non-relativistic field theoretical nuclear model of Johnson and
Teller\cite{ref:JoTe55}. The present form of QHD was established by
Chin and Walecka who reproduced the saturation
property of the nuclear matter within the mean field approximation in
the 70's\cite{ref:ChWa74}. Since then it has been
enjoying its success in accounting for various nuclear phenomena
\cite{ref:Ri96}. Presently it is appreciated as a reliable
way, alternative to traditional non-relativistic nuclear theories
such as the Skyrme-Hartree-Fock calculation, of describing not only
the ground state properties of finite(spherical\cite{ref:GaRiTh90},
deformed\cite{ref:PrWa87} and superdeformed\cite{ref:Ta96}) nuclei but
also the scattering observables\cite{ref:Co93}. Recently it has also
been applied extensively to exotic nuclei\cite{ref:Re96}.
Incorporating the polarization of the Fermi and Dirac sea, which is
neglected in the mean field approximation, on the other hand, QHD can
be an effective theory for the hadron properties such as masses of
vector mesons in finite density nuclear medium\cite{ref:HaShKu96} to
which lattice QCD calculations have not been available.

Two directions of extensions of the relativistic mean field theory to
the description of the excited states have been done so far. One is to
the giant resonances\cite{ref:NiKuSu87} and the other is to the yrast
states of rotating nuclei\cite{ref:Bi81,ref:KoRi89}. In
\cite{ref:KoRi89}, M\"unich group first applied this model to the
yrast states of $^{20}$Ne and got similar results to the
Skyrme-Hartree-Fock calculation. In some following
papers\cite{ref:KoeRi93}, they showed that this model could also
reproduce the moments of inertia of medium heavy and heavy superdeformed
nuclei in which effects of the pairing correlation was assumed to be
not important. This model was extended to the non-uniform
three-dimensional rotation and a method of its quantization was also
discussed\cite{ref:KaNaMa93}. From the theoretical point of view,
however, Koepf and Ring's formulation based on the Lorentz
transformation is not adequate because the rotating frame is an
accelerated one, and the coordinate transformation from the laboratory
frame to the rotating one is not a Lorentz but a general coordinate
transformation. The main reason why we adopt QHD, a special
relativistic model, is to respect the Lorentz covariance even if
velocities involved are not so large. Parallel to this, we should
adopt general relativistic models for the phenomena for which general
coordinate transformations are necessary even if the curvature of the
space-time is zero. Therefore, in this Letter, we develop a general
relativistic mean field theory for rotating nuclei by adopting the
tetrad formalism\cite{ref:We72}.

The crucial point is the transformation property of the nucleon
field under the coordinate transformation

\begin{eqnarray}
  x^{\alpha}=
  \left(
    \begin{array}{c}
      t \\
      \bbox{x}
    \end{array}
  \right)
  \rightarrow \tilde{x}^{\mu} & = &
  \left(
    \begin{array}{c}
      \tilde{t} \\
      \tilde{\bbox{x}}
    \end{array}
  \right)=
  \left(
    \begin{array}{cccc}
      1 & \bbox{0}^{T} \\
      \bbox{0} & R_{x}(t)
    \end{array}
  \right)
  \left(
    \begin{array}{c}
      t \\
      \bbox{x}
    \end{array}
  \right),
  \label{eqn:unirot} \\
  R_{x}(t) & = &
  \left(
    \begin{array}{ccc}
      1 & 0 & 0 \\
      0 & \cos\Omega t & \sin\Omega t \\
      0 & -\sin\Omega t & \cos\Omega t
    \end{array}
  \right),
\end{eqnarray}

\noindent
where $x^{\alpha}$ stands for the laboratory frame while $\tilde{x}
^{\mu}$ a frame uniformly rotating around the $\tilde{x}=x$ axis with
an angular velocity $\Omega$. Koepf and Ring adopted

\begin{equation}
  \psi(x)\rightarrow\tilde{\psi}(\tilde{x})
  =e^{i\Omega t\Sigma_{x}}\psi(x),\;
  \Sigma_{x}=\frac{\displaystyle 1}{\displaystyle 4}
  [\gamma^{2},\gamma^{3}],
  \label{eqn:spinor}
\end{equation}

\noindent
which is applicable only to the constant angle shift.
But obviously this is not for the present case. Therefore we have to adopt

\begin{equation}
  \psi(x)\rightarrow\tilde{\psi}(\tilde{x})=\psi (x),
  \eqnum{\ref{eqn:spinor}'}
\end{equation}

\noindent
as known in the quantum theory of gravity\cite{ref:We72}; the fermion
field transforms as a scalar under the general coordinate
transformation. The tetrad
formalism gives us how to treat the spinor field in general
relativity. In the following, we develop our formalism.

First we consider a non-inertial frame (either curved or flat)
represented by a metric tensor
$g_{\mu\nu}(\tilde{x})$. The principle of equivalence allows us to
construct a set of coordinates $\xi^{\alpha}_{X}(\tilde{x})$ that are
locally inertial at $\tilde{x}^{\mu}=X^{\mu}$. Then the metric tensors
of the non-inertial and the inertial frames are related as

\begin{equation}
  g_{\mu\nu}(\tilde{x})=V^{\alpha}_{\;\:\mu}(\tilde{x})
  V^{\beta}_{\;\:\nu}(\tilde{x})\eta_{\alpha\beta},
  \label{eqn:mettet}
\end{equation}

\noindent
here a tetrad is defined by

\begin{equation}
  V^{\alpha}_{\;\:\mu}(X)=\left(\frac{\partial\xi^{\alpha}
  _{X}(\tilde{x})} {\partial\tilde{x}^{\mu}}\right)
  _{\tilde{x}=X}.
  \label{eqn:tetrad}
\end{equation}

\noindent
Labels $\alpha,\beta,\cdots$ refer to the inertial frames while
$\mu,\nu,\cdots$ to the non-inertial ones.
This quantity, the tetrad, transforms
as a vector not only under the general coordinate transformation,

\begin{equation}
  V^{\alpha}_{\;\:\mu}(\tilde{x})\rightarrow
  V'^{\alpha}_{\;\:\mu}(\tilde{x}')=
  \frac{\partial\tilde{x}^{\nu}}{\partial\tilde{x}'^{\mu}}
  V^{\alpha}_{\;\:\nu}(\tilde{x}),
\end{equation}

\noindent
but also under the local Lorentz transformation,

\begin{equation}
  V^{\alpha}_{\;\:\mu}(X)\rightarrow
  V'^{\alpha}_{\;\:\mu}(X)=\Lambda^{\alpha}_{\;\:\beta}(X)
  V^{\beta}_{\;\:\mu}(X).
  \label{eqn:llt}
\end{equation}

\noindent
The latter which leaves eq.(\ref{eqn:mettet}) invariant allows us to
choose various forms for $V^{\alpha}_{\;\:\mu}$.
We will make use of this property later.
The main advantage of introducing the tetrad is that any tensors
$B^{\mu\nu\cdots}(\tilde{x})$ with respect to the general coordinate
transformation can be converted to scalars with respect to it, which is
at the same time tensors with respect to the local Lorentz
transformation,

\begin{eqnarray}
  & B^{\mu\nu\cdots} & (\tilde{x})\Longrightarrow
  ^{*}\!\!B^{\alpha\beta\cdots}(\tilde{x})=
  V^{\alpha}_{\;\:\mu}(\tilde{x})
  V^{\beta}_{\;\:\nu}(\tilde{x})\cdots B^{\mu\nu\cdots}
  (\tilde{x}) \\
  & : & \mbox{general coordinate scalar and Lorentz tensor,}
  \nonumber
\end{eqnarray}

\noindent
by contracting with the tetrad. This implies that this contraction is
not necessary for the general coordinate scalar quantities,

\begin{eqnarray}
  \phi(\tilde{x}) & \Longrightarrow & ^{*}\!\phi(\tilde{x})
  =\phi(\tilde{x}) \\
  & : & \mbox{general coordinate scalar and Lorentz scalar,}
  \nonumber \\
  \psi(\tilde{x}) & \Longrightarrow & ^{*}\!\psi(\tilde{x})
  =\psi(\tilde{x}) \\
  & : & \mbox{general coordinate scalar and Lorentz spinor.}
  \nonumber
\end{eqnarray}

\noindent
The covariant derivative with respect to the local Lorentz
transformation of the general coordinate scalar
$^{*}\!\varphi(=^{*}\!\!\psi,^{*}\!\phi,^{*}\!\!A^{\alpha},\cdots)$
is given by\cite{ref:We72}

\begin{equation}
  \tilde{\nabla}_{\alpha}\,^{*}\!\varphi=V_{\alpha}^{\:\mu}
  \tilde{\nabla}_{\mu}\,^{*}\!\varphi\equiv V_{\alpha}^{\:\mu}
  (\tilde{\partial}_{\mu}+\Gamma_{\mu})^{*}\!\varphi,
\end{equation}

\noindent
with the connection

\begin{equation}
  \Gamma_{\mu}(\tilde{x})=\frac{1}{2}\sigma^{\alpha\beta}
  V_{\alpha}^{\:\nu}(\tilde{x})V_{\beta\nu;\mu}(\tilde{x}),
  \label{eqn:FIco}
\end{equation}

\noindent
where $\sigma^{\alpha\beta}$ is the generator of the Lorentz group, the
symbol $;\mu$ denotes the well-known covariant derivative with respect
to the general coordinate transformation.

Collecting all the ingredients given above, we can generalize the
Lagrangian to the non-inertial frame with the following prescriptions:

\begin{enumerate}
  \renewcommand{\labelenumi}{\arabic{enumi})}
  \item
    Write the Lagrangian in the Minkowski space-time.
  \item
    Contract all the tensors with the tetrad.
  \item
    Replace all the derivatives with the covariant derivatives.
  \item
    Multiply $\sqrt{-g}$ to cast the resulting quantity into a scalar
    density with respect to the general coordinate transformation.
    Here
    \begin{equation}
      g={\rm det}(g_{\mu\nu}),\;
      \sqrt{-g}={\rm det}(V^{\alpha}_{\;\:\mu}).
    \end{equation}
\end{enumerate}

The results for the \sigome model,

\begin{equation}
  {\cal L}={\cal L}_{N}+{\cal L}_{\sigma}+{\cal L}_{\omega}
  +{\cal L}_{\rm int},
\end{equation}
\begin{eqnarray}
  {\cal L}_{N} & = & \overline{\psi}
    (i\gamma^{\alpha}\partial_{\alpha}-M)\psi, \\
  {\cal L}_{\sigma} & = & \frac{1}{2}(\partial_{\alpha}\sigma)
    (\partial^{\alpha}\sigma)-\frac{1}{2}m_{\sigma}^{2}\sigma^{2}, \\
  {\cal L}_{\omega} & = &
    -\frac{1}{4}F_{\alpha\beta}F^{\alpha\beta}
    +\frac{1}{2}m_{\omega}^{2}\omega_{\alpha}\omega^{\alpha}, \\
  {\cal L}_{\rm int} & = & g_{\sigma}\overline{\psi}\psi\sigma
    -g_{\omega}\overline{\psi}\gamma^{\alpha}\psi\omega_{\alpha},
\end{eqnarray}

\noindent
are

\begin{eqnarray}
  \lefteqn{{\cal L}_{N}\rightarrow} \nonumber \\
  & & \sqrt{-g}\left[\overline{\psi}(\tilde{x})
  (i\tilde{\gamma}^{\mu}(\tilde{x})
  (\tilde{\partial}_{\mu}+\Gamma_{\mu}(\tilde{x}))-M)
  \psi(\tilde{x})\right],
  \label{eqn:lagspinor}
\end{eqnarray}

\noindent
with eq.(\ref{eqn:FIco}),

\begin{eqnarray}
  {\cal L}_{\sigma} & \rightarrow & \nonumber \\
  & \sqrt{-g} & \left[\frac{\displaystyle 1}{\displaystyle 2}
  V_{\alpha}^{\:\mu}(\tilde{x})(\tilde{\nabla}_{\mu}
  \sigma(\tilde{x}))V^{\alpha}_{\;\:\nu}(\tilde{x})
  (\tilde{\nabla}^{\nu}\sigma(\tilde{x}))
  -\frac{\displaystyle 1}{\displaystyle 2}m_{\sigma}^{2}
  \sigma^{2}(\tilde{x})\right] \nonumber \\
  = & \sqrt{-g} & \left[\frac{\displaystyle 1}{\displaystyle 2}
  (\tilde{\partial}_{\mu}\sigma(\tilde{x}))
  (\tilde{\partial}^{\mu}\sigma(\tilde{x}))
  -\frac{\displaystyle 1}{\displaystyle 2}m_{\sigma}^{2}
  \sigma^{2}(\tilde{x})\right],
\end{eqnarray}

\noindent
owing to the fact that the covariant derivative coincides
with the ordinary one for the scalar field,

\begin{eqnarray}
  {\cal L}_{\omega} & \rightarrow & \nonumber \\
  & \sqrt{-g} & \left[-\frac{\displaystyle 1}{\displaystyle 4}
  \left(
    V_{\alpha}^{\:\mu}(\tilde{x})\tilde{\nabla}_{\mu}
    (V_{\beta}^{\:\nu}(\tilde{x})\omega_{\nu}(\tilde{x}))
    -V_{\beta}^{\:\mu}(\tilde{x})\tilde{\nabla}_{\mu}
    (V_{\alpha}^{\:\nu}(\tilde{x})\omega_{\nu}(\tilde{x}))
  \right)\right. \nonumber \\
  & & \times\left.\left(
    V^{\alpha}_{\;\:\mu}(\tilde{x})\tilde{\nabla}^{\mu}
    (V^{\beta}_{\;\:\nu}(\tilde{x})\omega^{\nu}(\tilde{x}))
    -V^{\beta}_{\;\:\mu}(\tilde{x})\tilde{\nabla}^{\mu}
    (V^{\alpha}_{\;\:\nu}(\tilde{x})\omega^{\nu}(\tilde{x}))
  \right)\right. \nonumber \\
  & & \left. +\frac{\displaystyle 1}{\displaystyle 2}
  m_{\omega}^{2}V_{\alpha}^{\:\mu}(\tilde{x})\omega_{\mu}(\tilde{x})
  V^{\alpha}_{\;\:\nu}(\tilde{x})\omega^{\nu}(\tilde{x})\right]
  \nonumber \\
  = & \sqrt{-g} & \left[-\frac{\displaystyle 1}{\displaystyle 4}
  F_{\mu\nu}(\tilde{x})F^{\mu\nu}(\tilde{x})
  +\frac{\displaystyle 1}{\displaystyle 2}m_{\omega}^{2}
  \omega_{\mu}(\tilde{x})\omega^{\mu}(\tilde{x})\right],
  \label{eqn:lagome}
\end{eqnarray}

\noindent
here

\begin{equation}
  F_{\mu\nu}(\tilde{x})=\omega_{\nu;\mu}(\tilde{x})
  -\omega_{\mu;\nu}(\tilde{x})
  =\tilde{\partial}_{\mu}\omega_{\nu}(\tilde{x})
  -\tilde{\partial}_{\nu}\omega_{\mu}(\tilde{x}),
\end{equation}

\noindent
and

\begin{eqnarray}
  \lefteqn{{\cal L}_{\rm int}\rightarrow} \nonumber \\
  & & \sqrt{-g}\left[g_{\sigma}\overline{\psi}(\tilde{x})
  \psi(\tilde{x})\sigma(\tilde{x})
  -g_{\omega}\overline{\psi}(\tilde{x})\tilde{\gamma}^{\mu}(\tilde{x})
  \psi(\tilde{x})\omega_{\mu}(\tilde{x})\right].
  \label{eqn:lagint}
\end{eqnarray}

\noindent
In (\ref{eqn:lagspinor}) and (\ref{eqn:lagint}), the generalized $\gamma$
matrices are defined as

\begin{equation}
  \tilde{\gamma}^{\mu}(\tilde{x})
  =\gamma^{\alpha}V_{\alpha}^{\:\mu}(\tilde{x}),
\end{equation}

\noindent
and they satisfy

\begin{equation}
  \left\{
    \tilde{\gamma}^{\mu}(\tilde{x}),\tilde{\gamma}^{\nu}(\tilde{x})
  \right\}
  =2g^{\mu\nu}(\tilde{x}).
\end{equation}

\noindent
The variational principle applied to the above generalized Lagrangian
for the non-inertial frame gives the equations of motion:

\begin{eqnarray}
  \left[i\tilde{\gamma}^{\mu}(\tilde{x})
  (\tilde{\partial}_{\mu}+\Gamma_{\mu}(\tilde{x}))-M
  +g_{\sigma}\sigma(\tilde{x}) \right. \nonumber \\
  \left. -g_{\omega}\tilde{\gamma}^{\mu}(\tilde{x})
  \omega_{\mu}(\tilde{x})\right]\psi(\tilde{x})=0,
\end{eqnarray}
\begin{equation}
  \tilde{\partial}_{\mu}\left[\tilde{\partial}^{\mu}
  \sigma(\tilde{x})\right]+m_{\sigma}^{2}\sigma(\tilde{x})=
  g_{\sigma}\overline{\psi}(\tilde{x})\psi(\tilde{x}),
\end{equation}

\noindent
and

\begin{equation}
  F^{\mu\nu}_{\;\;\;;\mu}(\tilde{x})+m_{\omega}\omega^{\nu}(\tilde{x})
  =g_{\omega}\overline{\psi}(\tilde{x})\tilde{\gamma}^{\nu}(\tilde{x})
  \psi(\tilde{x}),
\end{equation}

\noindent
for the nucleon, $\sigma$ meson, and $\omega$ meson, respectively.

Now we choose a specific flat but non-inertial frame,
that is a uniformly rotating frame given by the
coordinate transformation (\ref{eqn:unirot}). The metric tensor in
this case is

\begin{eqnarray}
  g_{\mu\nu}(\tilde{x}) & = & \frac{\partial x^{\alpha}}{\partial
  \tilde{x}^{\mu}}\frac{\partial x^{\beta}}{\partial\tilde{x}^{\nu}}
  \eta_{\alpha\beta} \nonumber \\
  & = &
  \left(
    \begin{array}{cccc}
      1-\Omega^{2}(\tilde{y}^{2}+\tilde{z}^{2}) & 0
        & \Omega\tilde{z} & -\Omega\tilde{y} \\
      0 & -1 & 0 & 0 \\
      \Omega\tilde{z} & 0 & -1 & 0 \\
      -\Omega\tilde{y} & 0 & 0 & -1
    \end{array}
  \right).
  \label{eqn:metric}
\end{eqnarray}

\noindent
Looking at eqs.(\ref{eqn:mettet}),(\ref{eqn:tetrad}), and
(\ref{eqn:metric}), the simplest choice of the tetrad is

\begin{eqnarray}
  V^{\alpha}_{\;\:\mu}(\tilde{x}) & = & \frac{\partial x^{\alpha}}
  {\partial\tilde{x}^{\mu}} \nonumber \\
  & = &
  \left(
    \begin{array}{cccc}
      1 & & \bbox{0}^{T} & \\
      0 & & & \\
      -\Omega(\tilde{y}\sin\Omega\tilde{t}
        +\tilde{z}\cos\Omega\tilde{t}) & & R_{x}^{T}(\tilde{t}) & \\
      \Omega(\tilde{y}\cos\Omega\tilde{t}
        -\tilde{z}\sin\Omega\tilde{t}) & & &
    \end{array}
  \right),
  \label{eqn:fundtet}
\end{eqnarray}

\noindent
i.e., the choice of the inertial coordinate

\begin{equation}
  \xi^{\alpha}_{X}(\tilde{x})=x^{\alpha}.
\end{equation}

\noindent
We call this the {\it fundamental choice}. This choice results in

\begin{equation}
  \Gamma_{\mu}(\tilde{x})=0,
  \label{eqn:fundcon}
\end{equation}

\noindent
for the spinor field, and accordingly

\begin{eqnarray}
  \left[
    (R_{x}(\tilde{t})\bbox{\alpha})\cdot(\frac{1}{i}
    \tilde{\bbox{\nabla}}-g_{\omega}\tilde{\bbox{\omega}}(\tilde{x}))
    +\beta(M-g_{\sigma}\sigma(\tilde{x}))\right. \nonumber \\
    \left. \raisebox{0pt}[3ex]{}
    +g_{\omega}\tilde{\omega}^{0}(\tilde{x})-\Omega\tilde{L}_{x}
  \right]
  \psi_{i}(\tilde{x})=i\tilde{\partial}_{0}\psi_{i}(\tilde{x}),
  \label{eqn:funddeq} \\
  \left[
    \tilde{\partial}_{0}^{2}-\tilde{\bbox{\nabla}}^{2}
    +m_{\sigma}^{2}-\Omega^{2}\tilde{L}_{x}^{2}
  \right]
  \sigma({\tilde{x}})=g_{\sigma}\rho_{s}(\tilde{x}),
  \label{eqn:fundkgsig} \\
  \left[
    \tilde{\partial}_{0}^{2}-\tilde{\bbox{\nabla}}^{2}
    +m_{\omega}^{2}-\Omega^{2}\tilde{L}_{x}^{2}
  \right]
  \tilde{\omega}^{0}({\tilde{x}})=g_{\omega}
  \tilde{\rho}_{v}(\tilde{x}),
  \label{eqn:fundkgomet} \\
  \left[
    \tilde{\partial}_{0}^{2}-\tilde{\bbox{\nabla}}^{2}
    +m_{\omega}^{2}-\Omega^{2}(\tilde{L}_{x}+S_{x})^{2}
  \right]
  \tilde{\bbox{\omega}}({\tilde{x}})=g_{\omega}
  \tilde{\bbox{\jmath}}_{v}(\tilde{x}),
  \label{eqn:fundkgomes}
\end{eqnarray}

\noindent
for the equations of motion within the mean field approximation where
the nucleon field $\psi$ is expanded in terms of single particle
states $\psi_{i}$s. Here
\begin{eqnarray}
  \rho_{s}(\tilde{x}) & = & \sum_{i}^{\rm occ}
    \overline{\psi}_{i}(\tilde{x})\psi_{i}(\tilde{x}), \\
  \rho_{v}(\tilde{x}) & = & \sum_{i}^{\rm occ}
    \psi_{i}^{\dagger}(\tilde{x})\psi_{i}(\tilde{x}), \\
  \bbox{j}_{v}(\tilde{x}) & = & \sum_{i}^{\rm occ}
    \psi_{i}^{\dagger}(\tilde{x})\bbox{\alpha}\psi_{i}(\tilde{x}),
\end{eqnarray}

\noindent
and a similar redefinition of the vector quantities according to Koepf
and Ring,

\begin{mathletters}
\begin{eqnarray}
  \tilde{\omega}^{0} & = & \omega^{0},\;\tilde{\bbox{\omega}}
  =\bbox{\omega}-(\bbox{\Omega}\times\tilde{\bbox{x}})\omega^{0},
  \label{eqn:fundredefome} \\
  \tilde{\rho}_{v} & = & \rho_{v},\;\tilde{\bbox{\jmath}}_{v}
  =R_{x}(\tilde{t})\bbox{j}_{v}-(\bbox{\Omega}\times
  \tilde{\bbox{x}})\rho_{v},
  \label{eqn:fundredefcur}
\end{eqnarray}
\end{mathletters}
was done. Note that $S_{x}$ is the ordinary spin operator for the spin=$1$
field,

\begin{equation}
  S_{x}=
  \left(
    \begin{array}{ccc}
      0 & 0 & 0 \\
      0 & 0 & -i \\
      0 & i & 0
    \end{array}
  \right).
\end{equation}

\noindent
Obviously eq.(\ref{eqn:funddeq}) is not stationary in the sense of the
ordinary cranking model because $R_{x}(\tilde{t})$ in the first term
is time dependent.

Therefore another choice of the tetrad is desirable to formulate a
stationary mean field theory parallel to the traditional
non-relativistic cranking model. This is possible by making use of the
degrees of freedom of the local Lorentz transformation (\ref{eqn:llt}).
This is due to the fact that $V^{\alpha}_{\;\:\mu}$ has 16 components
whereas only 10 components are independent. Utilizing this freedom of
choosing 6 components, we adopt another form for the tetrad,

\begin{equation}
  V^{\alpha}_{\;\:\mu}(\tilde{x})=
  \left(
    \begin{array}{cccc}
      1 & 0 & 0 & 0 \\
      0 & 1 & 0 & 0 \\
      -\Omega\tilde{z} & 0 & 1 & 0 \\
      \Omega\tilde{y} & 0 & 0 & 1
    \end{array}
  \right),
  \eqnum{\ref{eqn:fundtet}'}
\end{equation}

\noindent
which is called the {\it canonical choice}\cite{ref:OlTi62}. In
the present case, this corresponds to choosing the `instantaneously
rest frame' with respect to the rotating one as
the locally inertial frame. This choice results in

\begin{equation}
  \Gamma_{\mu}(\tilde{x})=
  \left(
    \begin{array}{c}
      -i\Omega\Sigma_{x} \\
      \bbox{0}
    \end{array}
  \right),
  \eqnum{\ref{eqn:fundcon}'}
\end{equation}

\noindent
for the spinor field, and accordingly

\begin{eqnarray}
  \left[
    \bbox{\alpha}\cdot(\frac{1}{i}\tilde{\bbox{\nabla}}-g_{\omega}
    \tilde{\bbox{\omega}}(\tilde{x}))+\beta(M-g_{\sigma}
    \sigma(\tilde{x}))\right. \nonumber \\
    \left. \raisebox{0pt}[3ex]{}
    +g_{\omega}\tilde{\omega}^{0}(\tilde{x})
    -\Omega(\tilde{L}_{x}+\Sigma_{x})
  \right]
  \psi_{i}(\tilde{x}) & = & i\tilde{\partial}_{0}\psi_{i}(\tilde{x}),
  \eqnum{\ref{eqn:funddeq}'} \\
  \left[
    \tilde{\partial}_{0}^{2}-\tilde{\bbox{\nabla}}^{2}
    +m_{\sigma}^{2}-\Omega^{2}\tilde{L}_{x}^{2}
  \right]
  \sigma({\tilde{x}}) & = & g_{\sigma}\rho_{s}(\tilde{x}),
  \eqnum{\ref{eqn:fundkgsig}'} \\
  \left[
    \tilde{\partial}_{0}^{2}-\tilde{\bbox{\nabla}}^{2}
    +m_{\omega}^{2}-\Omega^{2}\tilde{L}_{x}^{2}
  \right]
  \tilde{\omega}^{0}({\tilde{x}})
  & = & g_{\omega}\tilde{\rho}_{v}(\tilde{x}),
  \eqnum{\ref{eqn:fundkgomet}'} \\
  \left[
    \tilde{\partial}_{0}^{2}-\tilde{\bbox{\nabla}}^{2}
    +m_{\omega}^{2}-\Omega^{2}(\tilde{L}_{x}+S_{x})^{2}
  \right]
  \tilde{\bbox{\omega}}({\tilde{x}}) & = & g_{\omega}
  \tilde{\bbox{\jmath}}_{v}(\tilde{x}),
  \eqnum{\ref{eqn:fundkgomes}'}
\end{eqnarray}

\noindent
for the equations of motion. Eq.(\ref{eqn:fundredefcur}) is replaced by

%\begin{mathletters}
\begin{equation}
  \tilde{\rho}_{v}=\rho_{v},\;\tilde{\bbox{\jmath}}_{v}
  =\bbox{j}_{v}-(\bbox{\Omega}\times\tilde{\bbox{x}})\rho_{v},
  \eqnum{\ref{eqn:fundredefcur}'}
\end{equation}
%\end{mathletters}

\noindent
while (\ref{eqn:fundredefome}) is independent of the choice of the
tetrad. \\
Assuming the time dependence

\begin{equation}
  \psi_{i}(\tilde{t},\tilde{\bbox{x}})
    =\psi_{i}(\tilde{\bbox{x}})e^{-i\tilde{e}_{i}\tilde{t}},
\end{equation}

\noindent
where $\tilde{e}_{i}$ is the single-particle routhian, and the usual
$\tilde{t}$-independence of the meson mean fields, we come to the
desired stationary theory. This coincides with Koepf and Ring's.
The total energy in the laboratory frame, $\int d^{3}xT^{00}$,
can be calculated from the energy-momentum tensor in the
rotating frame,  $\tilde{T}^{\mu\nu}(\tilde{x})$, given by the
tetrad formalism\cite{ref:We72}. Again the result coincides with
theirs.

The reason why they obtained the correct expressions starting
from eq.(\ref{eqn:spinor}) is clear.
Since they defined the transformation property of the $\gamma$ matrices
such that $\overline{\psi} \gamma^\alpha \psi$ transforms as a (Lorentz)
vector, their transformed $\gamma$
matrices absorbed the inadequateness of the transformation property of
the fermion field. In addition, their transformation (\ref{eqn:spinor})
for the spinor and that for the $\gamma$ matrices
can be regarded as simulating the local Lorentz transformation from the
fundamental to the canonical tetrad. These implications are clarified
by constructing a correct general relativistic formulation.

To summarize, we have formulated a general relativistic mean field
theory for rotating nuclei adopting the tetrad formalism. We applied
this formulation to the \sigome model which has been known to give
good descriptions of various nuclear phenomena. We needed to adopt
the so-called canonical choice of the tetrad to obtain a stationary
equation of motion in the sense of the ordinary non-relativistic
cranking model. The results are the same as those of Koepf and Ring
who started from a special relativistic transformation property; their 
inadequateness was absorbed by their transformed $\gamma$
matrices\cite{ref:KoRi89}.

A possible way to go beyond the mean field approximation is the method
of Kaneko, Nakano and one of the present authors\cite{ref:KaNaMa93}
but any numerical application along this way has not been done. A
systematic numerical calculation of the yrast states of not only
stable but also unstable nuclei based on the present mean field theory 
is under progress and will be published separately.

We acknowledge the comments of K.Harada which inspired the present
study. Discussions with R.R.Hilton and J.K\"onig are also acknowledged.
This work was supported in part by the Grant-in-Aid for Scientific
Research from the Ministry of Education, Science and Culture
(No.08740209).

\end{document}